# A unified resource-pool architecture for high-dimensional direct-detection optical communication

Jingze Liu[1,†], Zhijuan Gu[1,†], Xinyang Yu[1,†], Ziwen Zhou[1], Zhuyixiao Liu[1], Mingming Zhang[1], Yuxuan Xiong[1], Peng Li[4], Zhongyao Luo[1], Jiajie Yuan[2], Hao Wu[1], Zhipei Sun[3], Siqi Yan[1,5,*], Yu Yu[1,*] and Ming Tang[1,2,5,*]

[1] School of Optical and Electronic Information and Wuhan National Laboratory for Optoelectronics, Huazhong University of Science and Technology, Wuhan 430074, China

[2] School of Future Technology, Huazhong University of Science and Technology, Wuhan 430074, China

[3] Department of Electronics and Nanoengineering, Aalto University, Espoo, Finland

[4] State Key Laboratory of Optical Fiber and Cable Manufacture Technology, Yangtze Optica Fibre and Cable Joint Stock Limited Company, Wuhan 430074, China

[5] Hubei Optical Fundamental Research Centre, Wuhan 430074, China

*Corresponding author: Siqi Yan, Email: siqya@hust.edu.cn; Yu Yu, Email: yuyu@mail.hust.edu.cn; Ming Tang, Email: tangming@hust.edu.cn

[†] These authors contributed equally to this work.

**Abstract**

Increasing optical communication capacity without proportionally increasing receiver complexity remains a key challenge for direct-detection links. Conventional systems typically assign wavelength, polarization and intensity to fixed, separately recovered functions, so that alphabet expansion is accompanied by additional demultiplexing, polarization handling, receiver branches and electronic processing. Here we introduce a unified resource-pool architecture for high-dimensional direct-detection optical communication, in which wavelength, polarization and intensity are jointly organized as a composite optical symbol space and recovered through optical-domain joint projection rather than dimension-by-dimension separation. The receiver is implemented with an integrated disordered photonic processor that transforms each composite optical state into a reproducible multi-output electrical fingerprint for single-shot direct recovery. In a dual-wavelength transmission experiment, the system resolves 4096 composite symbols, corresponding to 12 bits per symbol slot, with a bit error rate of $4.25 \times 10^{-4}$ after 10 km standard-fiber transmission. Additional experiments demonstrate dense polarization alphabets, wavelength-indexed state-space expansion and high-launch-power operation over hollow-core fiber. These results establish disorder-enabled joint projection in an integrated photonic processor as a route to hardware-efficient high-dimensional direct-detection communication beyond conventional dimension-partitioned receiver architecture.

## Introduction

The continued growth of data traffic is pushing optical communication systems towards higher capacity, higher spatial density and lower energy per bit under increasingly stringent constraints on hardware complexity and footprint[1–3]. Over the past decades, capacity scaling has relied on the progressive use of multiple physical degrees of freedom of light, including wavelength, polarization, spatial mode, and advanced modulation formats[4–7]. These advances have enabled modern short-reach interconnects, metro networks and long-haul transmission systems[8–11]. As optical systems approach practical limits in receiver complexity and power consumption, the key challenge is no longer only how many physical degrees of freedom can be accessed, but how efficiently they can be organized, detected, and converted into usable communication capacity.

Most optical communication architectures remain fundamentally dimension-partitioned. Wavelengths are treated as independently multiplexed channels, polarization is resolved into two orthogonal basis states, and intensity or field quadrature are recovered as separate signal variables[12–14]. This strategy is effective when different degrees of freedom can be cleanly separated and processed by dedicated receiver branches, but it ties capacity scaling to additional demultiplexing, polarization handling, optical filtering, high-speed electronics, or digital signal processing. The constraint is particularly severe in intensity-modulation direct-detection systems, which remain attractive because they avoid optical hybrids, local oscillators and coherent DSP, yet usually exploit multidimensional resources only in restricted forms: wavelength scaling adds filters and detector branches, polarization-division multiplexing is confined to two fixed channels, and intensity is detected as a scalar amplitude variable[15–18]. As a result, conventional direct-detection systems access only a limited portion of the available optical state space, creating a mismatch between the richness of light's physical degrees of freedom and the architecture used to recover information from them.

Recent progress in high-dimensional photodetection has shown that compact optical and optoelectronic structures can directly identify complex light fields carrying multiple degrees of freedom, including intensity, polarization, wavelength and spatial mode. Representative examples include geometric deep optical sensing, dispersion-assisted high-dimensional photodetection and fiber-tip multidimensional light identification[19–21]. These works establish high-dimensional light-field detection as an important direction in photonics. However, they have mainly been developed for optical-field characterization, imaging, sensing or pattern-recognition-type tasks. Their translation into a fiber-transmission optical communication receiver, where high-dimensional optical states serve as information-bearing symbols and are evaluated by communication-level metrics such as bit error rate, remains largely unexplored.

Here we introduce a unified resource-pool framework for high-dimensional optical communication. Instead of assigning wavelength, polarization, and intensity to fixed and separately recovered roles, we jointly organize them into a composite optical symbol space. Each transmitted symbol is defined by a specific combination of these physical variables, and the receiver is designed to recover the composite state directly rather than decomposing it into constituent dimensions. The distinction is important: the gain does not arise from introducing a new physical degree of freedom, but from reorganizing existing optical variables into a more expressive and directly detectable alphabet. Polarization, in particular, is used not only as a binary pair of orthogonal channels, but as a multi-state symbolic coordinate within the composite alphabet.

To implement this concept, we develop an integrated disordered photonic processor that functions as a compact high-dimensional direct detector. The processor deliberately mixes the wavelength, polarization, and intensity components of an incoming symbol through a fixed optical projection, producing a multi-output electrical fingerprint for each composite state. This converts multidimensional optical detection from explicit channel separation into passive optical-domain feature generation. A lightweight decision layer then recovers the transmitted symbol from the projected output features. In contrast to conventional receivers that suppress or avoid complex mixing, the present architecture uses integrated disorder as a resource for joint state projection, enabling enlarged optical alphabets to be detected without assigning a dedicated receiver chain to each physical dimension[22–24].

We experimentally validate the framework across multiple encoding strategies and transmission conditions. In a dual-wavelength implementation, the system resolves a 4096-state composite symbol space, formed by two wavelengths, four intensity levels and 16 polarization states, with an accuracy exceeding 99%. This corresponds to a 64-fold alphabet expansion relative to a conventional polarization-division-multiplexed direct-detection baseline. Additional experiments demonstrate dense polarization encoding, wavelength-indexed state-space expansion and fiber-link operation, including high-launch-power transmission over hollow-core fibre[25–27]. These results establish a hardware-efficient high-dimensional direct-detection architecture beyond conventional dimension-partitioned optical communication systems.

## Results

### Unified resource-pool framework and integrated joint projection

The central objective of a communication architecture is to maximize the recoverable information between the transmitted symbol and the received observation under practical constraints on hardware complexity (HC), footprint and energy consumption[28,29]. For a transmitted symbol S and receiver observation Y, this recoverable information is governed by the mutual information

$$R = I(S;Y) = H(S) - H(S \mid Y) \tag{1}$$

where H(S) represents the entropy of the transmitted symbol alphabet and H(S|Y) denotes the residual uncertainty after transmission and detection[30]. This formulation provides a direct basis for evaluating multidimensional optical communication: capacity scaling requires not only increasing the size of the transmitted alphabet, but also ensuring that the enlarged symbol set remains distinguishable at the receiver[31].

Conventional IMDD architectures increase H(S) mainly through dimension-partitioned scaling. As illustrated in Fig. 1(a), wavelength, polarization and intensity are assigned predefined and separately processed roles: wavelengths form independently multiplexed channels, polarization is typically resolved into two orthogonal states, and intensity is modulated and detected as a scalar amplitude variable[32,33]. Although effective, this architecture ties alphabet expansion to additional demultiplexing, polarization handling, detector branches and sampling channels[34,35]. As the symbol space grows, receiver complexity increases because the system must preserve channel separation across each physical dimension[29].

The unified resource-pool framework proposed here follows a different principle. Instead of treating wavelength, polarization and intensity as isolated channels or functions, each transmitted symbol is defined as a composite optical state,

$$S = (\Lambda_i, P_j, I_k) \in \Lambda \times P \times I \tag{2}$$

where Λ, P and I represent the discrete wavelength, polarization, and intensity states, respectively. The framework therefore increases H(S) by forming a joint wavelength–polarization–intensity alphabet, while the receiver is designed to reduce H(S|Y) through direct recovery of the composite symbol rather than dimension-by-dimension decomposition. The key requirement is consequently not the separate extraction of each optical variable, but the preservation of distinguishable composite states after photonic projection.

To realize this requirement, we design and fabricate an integrated disordered photonic processor that maps each incoming composite state onto a directly distinguishable multi-output feature vector. The processor performs a fixed optical projection described by

$$\mathbf{Y_S} = D(\mathbf{H}\mathbf{X_S}) + \mathbf{n} \tag{3}$$

where $\mathbf{X_S}$ represents the optical field associated with symbol $S$, $\mathbf{H}$ is the transfer matrix of the disordered photonic processor, $D(\cdot)$ denotes direct photodetection and $\mathbf{n}$ accounts for system noise. Physically, the direct-detection $D(\cdot)$ acts as a quadratic square-law transient, transforming the mixed optical field into an electrical intensity vector $\mathbf{Y_S}$ that inherently comprises both squared self-terms and pairwise cross-interference terms of the coupled modes. Reliable recovery is therefore determined by the separation between projected feature vectors rather than by independent recovery of wavelength, polarization, and intensity. The integrated disorder provides spatial diversity and decorrelation, converting high-dimensional optical-state discrimination into a compact feature-generation process at the photonic front end[24]. The relationship between feature-space separation, minimum Euclidean distance and symbol recovery probability is derived in **Supplementary Note 1**.

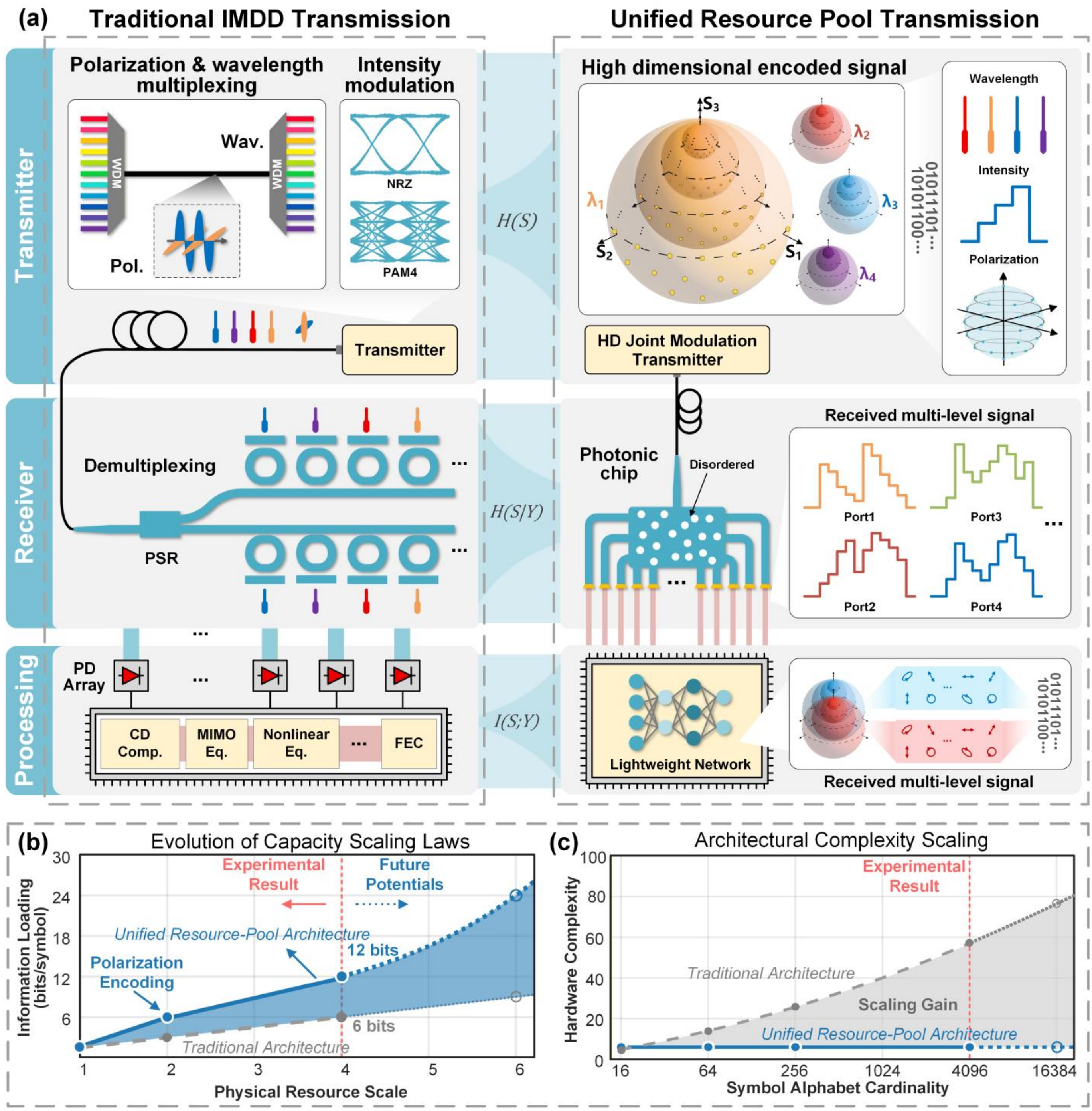

**Figure 1: Unified resource pool framework for high-dimensional optical communication.** (a) Structural comparison between traditional dimension-partitioned and unified resource-pool architectures illustrating the transition from isolated channels to a cohesive state-space paradigm. (b) Comparative scaling of information loading per symbol relative to the physical resource scale. The unified architecture achieves accelerated capacity gains by exploiting multidimensional synergies compared to the linear growth of conventional systems. (c) Evolution of receiver hardware complexity demonstrating the fundamental decoupling of symbol cardinality from implementation overhead realized through the integrated joint-projection engine.

The resulting scaling advantage is quantified in Fig. 1(b) by comparing information loading (IL) with the normalized physical resource scale (PRS). Here the physical resource scale Γ denotes the number of independently accessed optical resources, such as wavelength channels and polarization-state occupancy, whereas information loading denotes the net mutual information recovered per symbol slot. The detailed calculation procedure for these two quantities is provided in **Supplementary Note 2**. In a dimension-partitioned architecture, information loading increases additively as independent resources are added [8,34]. In the unified resource-pool architecture, the joint alphabet expands multiplicatively because each symbol is selected from the Cartesian product of wavelength, polarization, and intensity states. At the experimental scale of two wavelengths, four intensity levels and 16 polarization states per wavelength, the framework increases the information loading from 6 bits to 12 bits per symbol slot, corresponding to a 64-fold expansion of the usable symbol alphabet relative to a conventional polarization-division-multiplexed direct-detection baseline.

This expansion is achieved without a proportional increase in receiver hardware. Figure 1c compares the normalized receiver complexity as a function of symbol alphabet cardinality, where the complexity metric accounts for active photodetectors, sampling channels and discrete optical demultiplexing units, as defined in **Supplementary Note 2**. Conventional dimension-partitioned scaling requires additional hardware to maintain channel separation as the alphabet grows [35]. In contrast, the proposed architecture maintains a compact receiver footprint because high-cardinality discrimination is delegated to the fixed optical projection of the integrated disordered processor. The shaded region in Fig. 1(c) therefore identifies the regime in which composite alphabet expansion is converted into increased information loading without equivalent receiver-side hardware escalation. This decoupling of symbol-space growth from dimension-by-dimension receiver scaling is the central architectural advantage of the unified resource-pool framework.

To provide a rigorous theoretical foundation for these scaling laws, we benchmark the quantitative scaling performance of both architectures across symbol alphabet cardinalities ranging from $M$ = 16 to 16384 as summarized in Table 1 Under a conventional PDM-PAM4 baseline where each wavelength channel carries 3 bits of net capacity, resolving higher-dimensional cardinalities necessitates a proportional escalation of both PRS and HC. At a moderate cardinality of M = 64, which is equivalent to 6 bits per symbol, the proposed framework saves 50% of the PRS by reducing the scale from $\Gamma_{\mathrm{trad}} = 4$ to $\Gamma_{\mathrm{prop}} = 2$, and reduces HC by 52.6% by decreasing the complexity from $C_{\mathrm{trad}} = 19$ to $C_{\mathrm{prop}} = 9$. To reach the experimentally demonstrated 12-bit capacity at M = 4096, the traditional baseline requires a PRS of 8, yielding a HC of $C$ = 53. In stark contrast, the proposed unified resource-pool framework delivers the identical 12-bit capacity at a constant PRS of 4 and a HC of 9. This achieves a 50% reduction in the occupied PRS alongside an 83% reduction in HC. Extrapolating to the 14-bit capacity at $M$ = 16384, the resource-pool framework is envisioned to maintain its hardware footprint of $C = 9$ at $\Gamma = 4$, widening the complexity savings to 88.2% and physical resource savings to 60%. These quantitative benchmarks substantiate the physical-layer decoupling of capacity growth from hardware escalation, establishing integrated disorder-enabled joint projection as a highly efficient path for high-dimensional optical communication.

**Table 1 | Quantitative comparison of scaling efficiency under different paradigms**

| Cardinality (M) | IL (bits/symbol) | Trad. WDM (PRS/HC) [a] | Prop. Resource Pool (PRS/HC) | • Quantitative Scaling Gain • Comparison [b] |
|---|---|---|---|---|
| 16 | 4 | $\Gamma = 2 / C = 8$ | $\Gamma = 2 / C = 9$ | • Constant Physical Resource • +12.5% Hardware Complexity |
| 64 | 6 | $\Gamma = 4 / C = 19$ | $\Gamma = 2 / C = 9$ | • -50% Physical Resource • -52.6% Hardware Complexity |
| 256 | 8 | $\Gamma = 6 / C = 34$ | $\Gamma = 4 / C = 9$ | • -33% Physical Resource • -73.5% Hardware Complexity |
| 4096 [c] | 12 | $\Gamma = 8 / C = 53$ | $\Gamma = 4 / C = 9$ | • -50% Physical Resource • -83.0% Hardware Complexity |
| 16384 [d] | 14 | $\Gamma = 10 / C = 76$ | $\Gamma = 4 / C = 9$ | • -60% Physical Resource • -88.2% Hardware Complexity |

[a] Under the conventional PDM-PAM4 baseline, each wavelength channel carries 2 polarizations with PAM-4, effectively yielding 3 bits/symbol due to direct-detection crosstalk.

[b] The relative physical resource saving and hardware complexity reduction are calculated as $1 - \Gamma_{prop} / \Gamma_{trad}$ and $1 - C_{prop} / C_{trad}$, respectively.

[c] Experimental demonstration point.

[d] Extrapolated potential point.

**Integrated disordered photonic processor for joint projection of composite optical states**

The composite-state projection engine is physically realized as an integrated disordered photonic processor fabricated on a silicon-on-insulator platform. This device serves as the photonic front end of the unified resource pool architecture by converting multidimensional composite optical states into distinguishable multi-port electrical features for direct recovery. The microscope image in Figure 2(b) displays the monolithic integration of the entire photonic chip which comprises a disordered scattering region coupled to a sixteen-channel photodetector array. A magnified view of the disordered scattering structure is presented in Figure 2(a) while Figure 2(c) illustrates the layout of the on-chip photodetector units. Unlike conventional receivers that resolve wavelength and polarization through discrete filters and splitters, the processor performs joint optical projection within a single integrated structure. Its operation relies on disorder-induced multipath interference which extends the effective optical path length within a micron-scale footprint and generates a high-dimensional response that is jointly sensitive to wavelength and polarization. The theoretical stability of the resulting transfer matrix **H** is analyzed in the **Supplementary Note 3**.

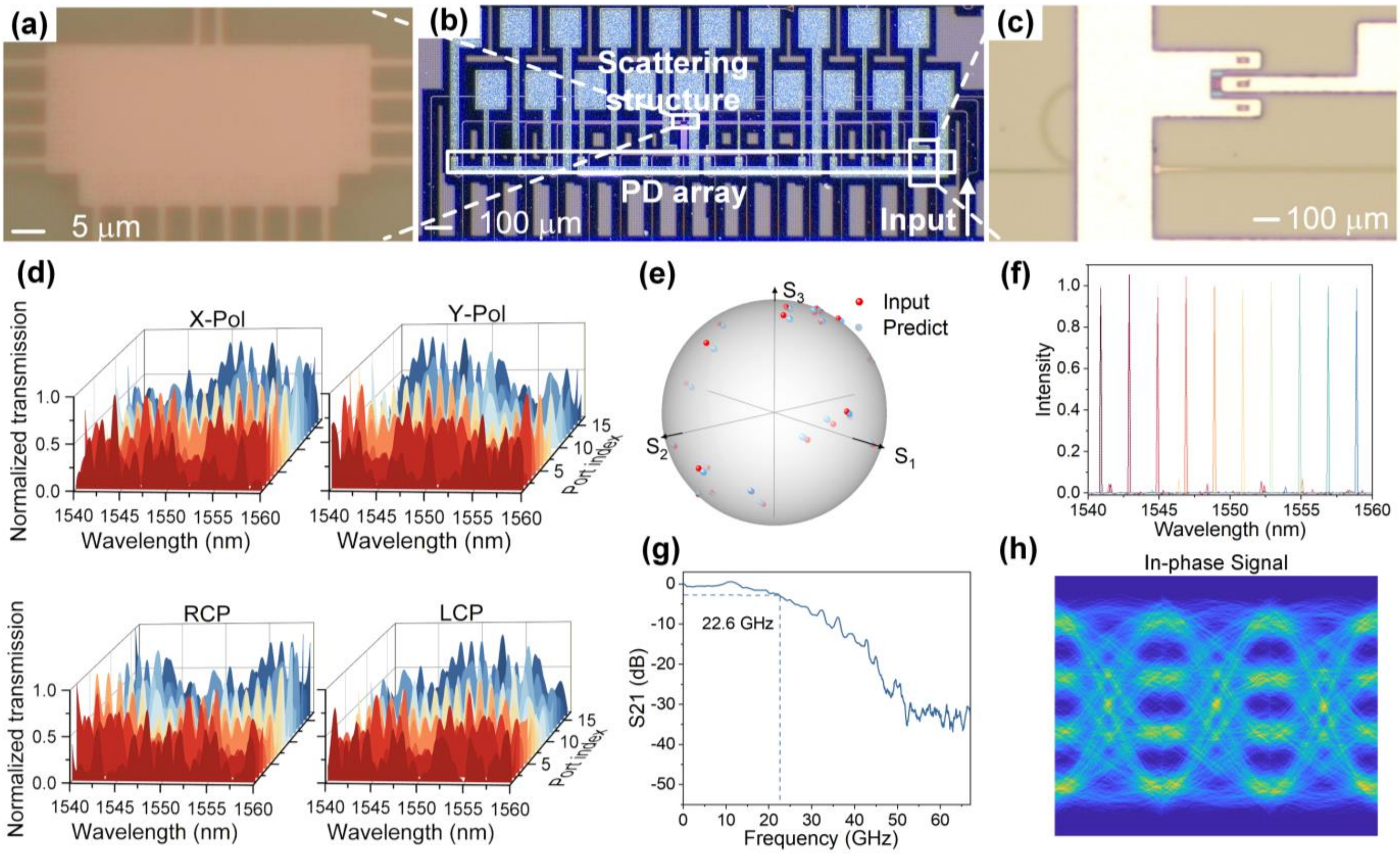

**Figure 2 | Integrated disordered photonic processor for projection of composite optical states.** (a) Magnified optical microscope image of the disordered scattering structure. (b) Optical microscope image of the fabricated silicon photonic chip integrating the scattering region and photodetector array. (c) Microscope image of the sixteen-channel photodetector array. (d) Normalized multi-port transmission spectra for horizontal, vertical, right-hand circular and left-hand circular input polarization states from 1540 nm to 1560 nm. (e) Experimental reconstruction of polarization states on the Poincare sphere, showing a polarization sensing precision of 1.2 degrees. (f) Spectral discrimination performance with a decorrelation bandwidth of 0.4 nm. (g) Measured optoelectronic frequency response of the integrated photodetectors, showing a 3-dB bandwidth of 22.6 GHz. (h) Experimental four-level pulse-amplitude-modulation eye diagram measured through the integrated processor.

We first characterize the multidimensional response of the processor across the spectral and polarization domains. Figure 2d shows representative normalized multi-port response spectra for horizontal, vertical, right-hand circular and left-hand circular polarization states over the wavelength range from 1540 nm to 1560 nm. The output patterns exhibit a strong and reproducible dependence on both input variables, confirming that the processor maps the multidimensional input space into a diverse detector-observation space. This behavior is a prerequisite for resource-pool communication because the massive state-space cardinality can only be resolved if the constituent optical states remain sufficiently separated after photonic projection.

The quantitative accuracy of the processor is further evaluated through the reconstruction of individual optical coordinates. Figure 2(e) illustrates the experimental reconstruction of polarization states distributed across the Poincaré sphere where a sensing precision of 1.2 degrees is achieved. This high angular resolution supports the use of polarization beyond a binary orthogonal basis, allowing neighboring polarization states to remain distinguishable when incorporated into a larger composite alphabet. The wavelength response is characterized in Figure 2(f) where distinct spectral features are resolved with a decorrelation bandwidth of 0.4 nm. This spectral selectivity provides a critical discriminating dimension when wavelength is included in the unified state-space.

We further evaluate the dynamic capability of the integrated processor. As shown in Fig. 2(g), the integrated photodetectors exhibit a 3-dB optoelectronic bandwidth of 22.6 GHz. A clear four-level pulse-amplitude-modulation eye diagram is obtained in Fig. 2(h), confirming that the photodetection path itself supports high-speed temporal response. The present system-level composite-symbol experiment is operated at 100 MHz because the current package and multi-channel electrical readout were not designed as a fully RF-optimized communication module. Thus, the measurements establish the bandwidth potential of the photonic front end while leaving full high-baud resource-pool transmission to a dedicated high-speed implementation.

**Experimental construction and recovery of enlarged composite symbol spaces**

The unified resource-pool framework is experimentally implemented using a packaged photonic–electronic prototype. As shown in Fig. 3(a) the transmitter (Tx) constructs a multidimensional resource pool in which information is jointly encoded across wavelength, intensity, and polarization. In the dual-wavelength configuration, two optical carriers at 1550 and 1551 nm are used simultaneously. Each wavelength carries an independent 64-state sub-symbol formed by four intensity levels and 16 polarization states, as illustrated in Fig. 3(b). The joint two-wavelength state therefore forms a $64\times64 = 4096$-state composite alphabet, corresponding to 12 bits per symbol slot. This construction is central to the proposed architecture: the information is encoded in the composite optical state rather than in separately demultiplexed wavelength or polarization channels. Detailed hardware specifications are provided in the **Methods**, and the complete experimental setup is shown in **Supplementary Note 4**.

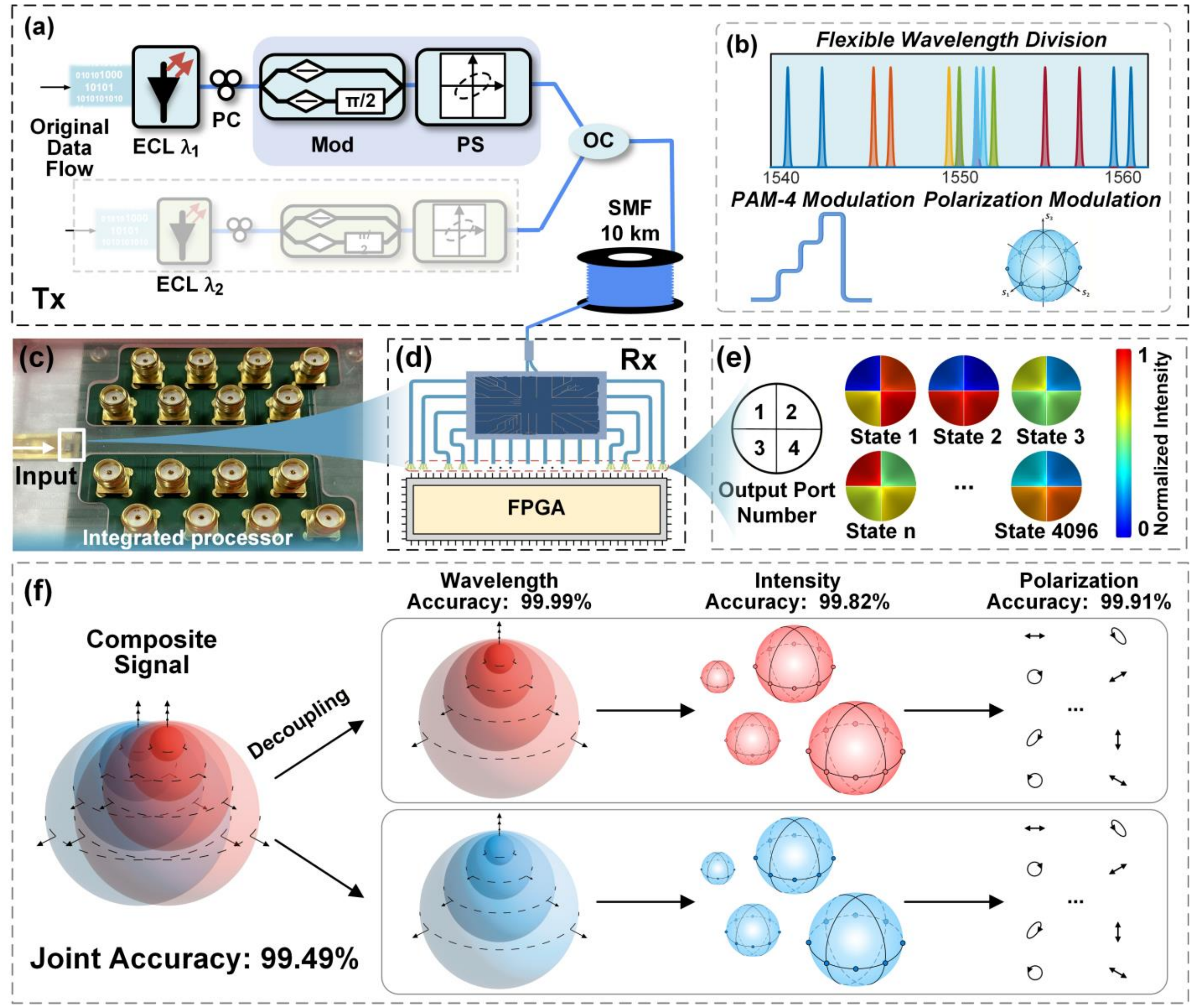

**Figure 3 | Construction and experimental recovery of enlarged composite symbol spaces.** (a) Schematic of the Tx architecture. (b) Insets illustrating the flexible wavelength division and the combined PAM-4 intensity and dense polarization modulation schemes. (c) Photograph of the physically packaged integrated processor. (d) Artistic schematic of the Rx chip. (e) Measured 4-port intensity responses serving as distinct spatial fingerprints for the 4096-state composite alphabet. (f) Hierarchical decoding results, demonstrating a joint recovery accuracy of 99.49% and high accuracy across individual physical dimensions.

At the receiver (Rx), the composite optical states are processed by the integrated disordered photonic processor, as schematically shown in Fig. 3(d). The packaged receiver module is shown in Fig. 3(c). In the communication experiment, four detector outputs are used for recovery, emphasizing that the enlarged alphabet is identified through joint photonic projection rather than through a dedicated receiver branch for each wavelength, polarization, or intensity state. The incident composite optical states are mapped by the scattering region onto deterministic four-port intensity fingerprints. Figure 3(e) presents representative measured four-port output distributions across the 4096-state composite alphabet. Each state produces a distinct projected feature vector, which serves as the observation space for digital symbol recovery. The data-acquisition procedure and normalization method are described in the Methods, and the complete set of four-port responses for all 4096 states is provided in **Supplementary Note 5**.

The recovery performance is quantified in Fig. 3(f). The classifier is trained using calibrated projected feature vectors and evaluated on independent test measurements. After transmission over a 10-km fiber link at a received optical power of −7 dBm, the system achieves a joint composite-state recovery accuracy of 99.49% for the full 4096-state alphabet. This corresponds to reliable recovery of 12 bits per symbol slot and a 64-fold alphabet expansion relative to a conventional polarization-division-multiplexed direct-detection baseline. A dimension-resolved analysis after joint recovery further shows that wavelength, intensity, and polarization are each identified with accuracies over 99%, confirming that the joint-projection receiver preserves the distinguishability of the constituent variables within the enlarged composite symbol space. The complete classification results, including confusion matrices for the individual dimensions and the joint alphabet, are provided in **Supplementary Note 6**.

**Transmission validation across various fiber links**

Having established composite-state recovery in the packaged prototype, we next use fiber links as controlled perturbation channels to examine the robustness of the projected feature space against loss, transmission reach and launch-power-induced nonlinearities. To establish a rigorous connection between the physical projection space and the digital transmission layer, we implement a structured bit-to-symbol mapping for the 4096-state, 12-bit composite alphabet, conceptually demonstrated in Fig. 4(a). Specifically, each 12-bit word is divided into two 6-bit sub-symbols carried by the two wavelength components, where the encoding schemes and signal waveforms within these dual-wavelength channels are conceptually mapped. Within each wavelength component, two bits define the four-level intensity state and four bits define one of 16 polarization states using a Gray-like labeling scheme[36]. These dual-channel signals are launched into a variable fiber link, combining into a high-dimensional composite symbol within the fiber before undergoing joint photonic projection at the receiver. The predicted state index is then mapped back to the corresponding 12-bit word using an inverse lookup table, enabling both the symbol error rate (SER) and the bit error rate (BER) to be calculated from independent test measurements. The training and testing datasets are strictly partitioned: the decoding model is established only from the calibration dataset, while independent measurements are reserved for validation.

Because the physical dimensions of wavelength, intensity, and polarization are orthogonal, the combination preserves the Gray-like characteristics, meaning that adjacent physical states with a high likelihood of confusion have a minimum Hamming distance of one. Under this multi-dimensional Gray-like mapping at medium-to-high signal-to-noise ratios, a single symbol error typically results in only a single bit error, allowing the BER to be closely approximated as [8]:

$$BER \approx SER / \log_2(M) = (1 - Accuracy) / 12$$

where M = 4096 is the cardinality of the composite alphabet. This mathematical formulation establishes the benchmarks for error correction: the standard 7% hard-decision FEC (HD-FEC) limit (BER = $3.8 \times 10^{-3}$) corresponds to a joint recovery accuracy threshold of 95.44%, while the 20% soft-decision FEC (SD-FEC) limit (BER = $2.0 \times 10^{-2}$) corresponds to a threshold of 76.0%.

We first characterize the receiver sensitivity of the resource-pool architecture over a 10-km standard single-mode fiber (SMF) link. As shown in Fig. 4(b), the received optical power is varied from -5 dBm to -9 dBm to evaluate operation close to the receiver noise floor. The joint composite-state recovery accuracy remains high across this range, decreasing from near-perfect recovery to 98.12% at -9 dBm. Based on our mapping, this lowest accuracy corresponds to a BER of approximately $1.57 \times 10^{-3}$, which is comfortably below the 7% HD-FEC threshold. At the reference operating point of -7 dBm, the dual-wavelength implementation achieves a joint recovery accuracy of 99.49% (BER ≈ $4.25 \times 10^{-4}$). We then evaluate the transmission reach in standard fiber, as shown in Fig. 4(c). The projected fingerprints remain distinguishable over increasing link lengths, with the full 4096-state alphabet recovered with an accuracy of 97.12% (BER ≈ $2.4 \times 10^{-3}$) after 30 km, maintaining its error-correctable status safely above the 7% HD-FEC limit.

Longer fiber links introduce cumulative attenuation, chromatic dispersion, and dynamic polarization drift, which distort the physical properties of the composite optical states and eventually compromise the separability of the projected fingerprints. Increasing the launch power is a direct way to restore the power budget, but in standard silica fiber it also introduces nonlinear and back-scattering impairments that can distort the composite optical state. We therefore examine whether the resource-pool alphabet remains recoverable in the high-launch-power regime.

In standard SMF, the joint recovery performance degrades markedly at elevated launch powers (Fig. 4(e)). The joint BER crosses above the 7% HD-FEC threshold when the launch power exceeds 12 dBm and falls outside the 20% soft-decision FEC limit beyond 16 dBm. Back-scattered-power measurements in **Supplementary Note 7** identify stimulated Brillouin scattering as the primary impairment in this regime[37].

Dimension-resolved analysis (Fig. 4(f, g)) reveals how these link impairments affect different constituent variables of the composite state. Specifically, the degradation is heavily dominated by the intensity dimension (Fig. 4(f)), where the SMF intensity BER rapidly rises to $9.57 \times 10^{-2}$ at 17 dBm. In contrast, the polarization-state identification (Fig. 4(g)) remains comparatively stable, staying below $3.0 \times 10^{-3}$ even at the highest launch power. This analysis shows that the joint-projection receiver can expose dimension-specific impairments, although recovery of the full alphabet is ultimately limited when one coordinate is strongly distorted.

To assess whether this limitation can be relaxed in a lower-nonlinearity transmission medium, we repeat the high-power experiment over a 24.3-km hollow-core fiber (HCF) link, the cross-sectional structure of which is shown in Fig. 4(d). Hollow-core fiber reduces the light–glass interaction and suppresses nonlinear impairments compared with conventional silica-core fiber, making it well suited for high-launch-power transmission[38,39]. In contrast to the SMF link, the HCF measurements show stable joint, intensity, and polarization BERs over the measured launch-power range up to 17 dBm (Fig. 4(e-g)). The joint BER (Fig. 4(e)) stays below $1.75 \times 10^{-3}$ even at a high launch power of 17 dBm, which is comfortably within the 7% HD-FEC limit. Strikingly, by suppressing nonlinear distortion, the HCF link bypasses the need for high-complexity 20% SD-FEC entirely. This result indicates that low-nonlinearity transmission media can preserve the projected feature space under launch powers that degrade standard fiber. Based on the measured receiver threshold and link power budget, the HCF platform suggests a route towards 100-km-class single-span operation while maintaining recovery of the 12-bit composite alphabet. The HCF experiment therefore serves as a stringent link-level test of the resource-pool architecture rather than as a prerequisite for its operation.

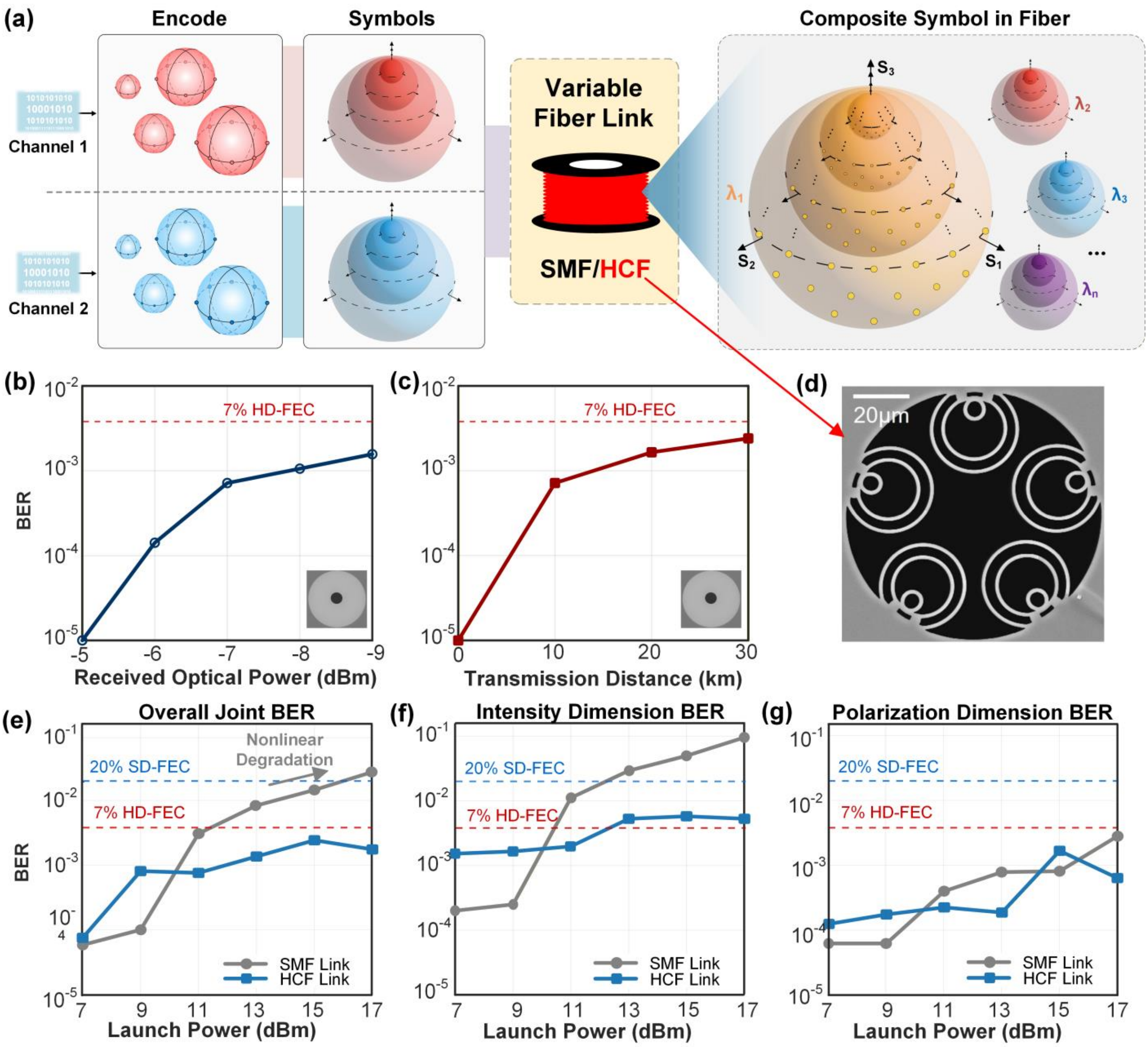

**Figure 4 | Transmission validation across fiber links.** (a) Conceptual demonstration of the experimental transmission link, illustrating the bitstream mapping, dual-channel encoding with wavelength components, propagation through the fiber link, and the resultant high-dimensional composite symbol within the fiber. (b) Power sensitivity analysis depicting the measured BER as a function of the received optical power, demonstrating operation relative to the 7% hard-decision HD-FEC limit. (c) Distance robustness representing the BER across transmission lengths up to 30 km. The insets in (b-c) shows a scanning electron micrograph (SEM) of the employed SMF cross-section. (d) Cross-sectional SEM image of the employed HCF. (e–g) Evaluation of nonlinear degradation under high launch powers comparing standard SMF (grey) and low-nonlinearity HCF (blue) links. The comparison is decoupled and plotted in terms of the overall joint BER (e), the intensity dimension BER (f), and the polarization dimension BER (g). The horizontal dashed lines represent the 7% HD-FEC (red) and 20% soft-decision FEC (SD-FEC, blue) limits, highlighting the dimension-specific penalties in SMF and the sustained stability of HCF.

## Dimensional expansion of the composite symbol space

To further assess the scalability of the unified resource-pool framework, we examine two additional expansion routes beyond the representative 4096-state dual-wavelength demonstration, conceptually illustrated in Fig. 5(a). These experiments test whether the same joint-projection principle can accommodate denser polarization alphabets and larger wavelength-indexed state spaces without reverting to dimension-by-dimension optical separation. In both cases, the receiver remains the same passive integrated photonic processor with a compact multi-port detection interface.

We first investigate polarization-state densification. Motivated by the observed robustness of polarization recovery under fiber transmission and elevated launch powers, we construct a compressed single-wavelength resource pool comprising 32 discrete polarization states and four intensity levels. This yields a 128-state composite alphabet (PAM4-32Pol) in which neighboring polarization states occupy a much denser region of the Jones space than in the 16-state configuration. Conventional recovery of such fine polarization variations would typically require explicit polarization analysis or Stokes-vector reconstruction with dedicated receiver hardware [32]. In contrast, the integrated disordered photonic processor directly maps these dense polarization–intensity states into distinguishable output fingerprints. As shown in the recognition-fidelity heatmap in Fig. 5(b), the results confirm accurate recovery across the full 128-state (L1 to L4 intensity levels across 32 polarization indices) alphabet, with most symbols reaching 100% fidelity and the minimum symbol accuracy remaining above 99%. These results show that polarization can serve as a dense symbolic coordinate within the composite optical state space, rather than only as a binary multiplexing dimension.

We next examine expansion along the wavelength dimension. Unlike conventional WDM, where different wavelengths are treated as independent parallel channels, here wavelength is incorporated as an encoding variable within the composite optical state. We increase the number of wavelength states from 1 to 16, spanning a spectral range from 1542 to 1557 nm with a channel spacing of 1 nm, and combine them with 16 polarization states and four intensity levels, producing a maximum alphabet cardinality of $16\times16\times4=1024$ states, corresponding to 10 bits per symbol. As summarized in Fig. 5(c), the recovered information loading increases logarithmically from 6 to 10 bits/symbol with the number of wavelength states, while the joint classification accuracy remains above 99.1% even at the 1024-state configuration. This result confirms that the processor provides sufficient wavelength-dependent decorrelation to support spectral-state expansion within the same joint-projection receiver architecture.

Together, these experiments show that the resource-pool framework is not restricted to a single alphabet design. Dense polarization encoding increases the number of usable states within a fixed wavelength channel, while wavelength-indexed encoding introduces an additional scalable degree of freedom for enlarging the composite alphabet. In both cases, the integrated disordered photonic processor converts expanded wavelength–polarization–intensity state spaces into directly recoverable output features, supporting high-dimensional IMDD communication without reverting to conventional dimension-partitioned receiver scaling.

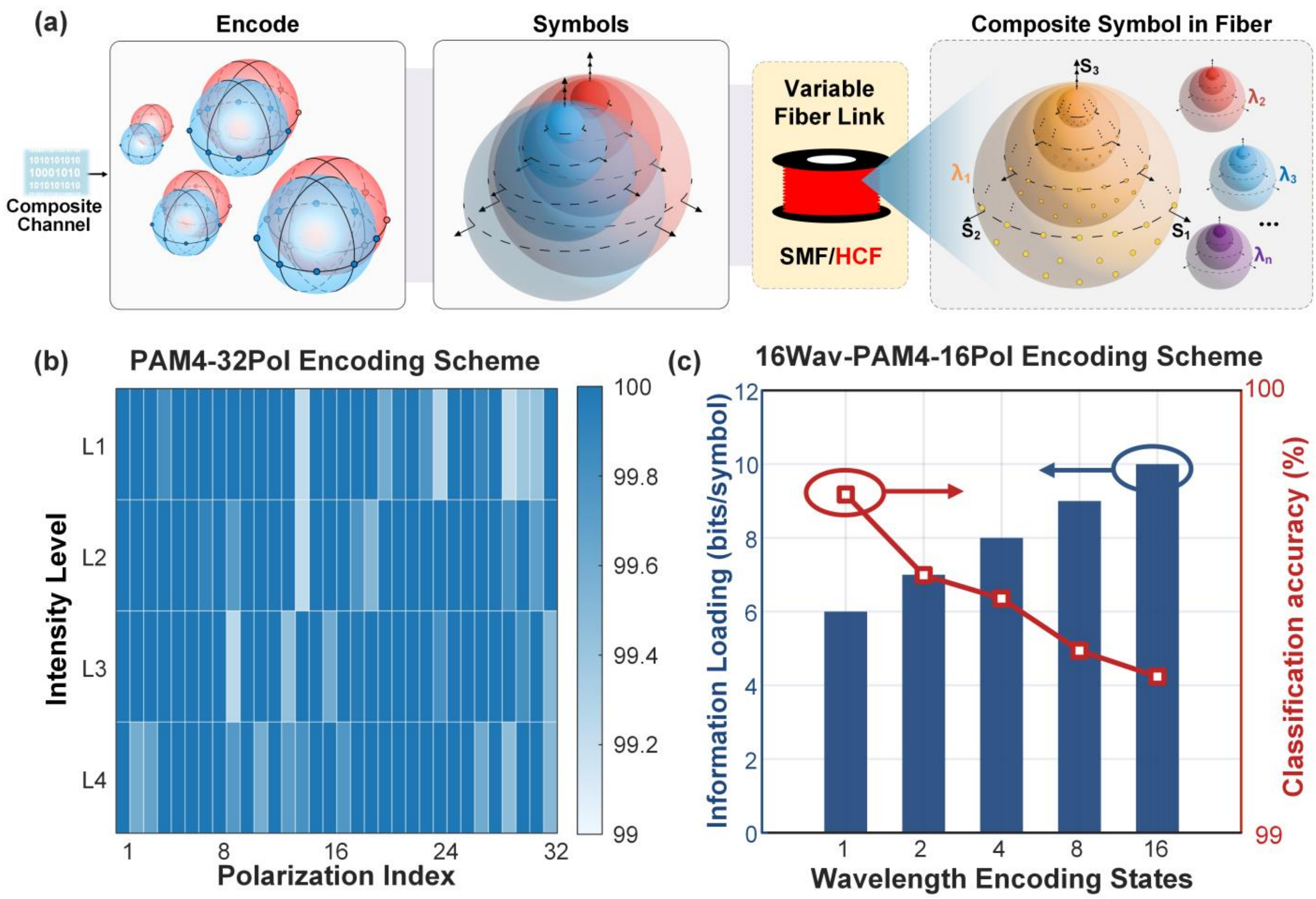

**Figure 5 | Dimensional expansion and scalability of the composite symbol space.** (a) Schematic of composite-symbol expansion through polarization densification and wavelength indexing. (b) Recognition accuracy heatmap for a single-wavelength PAM4-32Pol alphabet. (c) Information loading and joint classification accuracy as the wavelength encoding coordinate is scaled using PAM4 intensity and 16 polarization states.

## Discussion

The results reported here establish a system-level route to high-dimensional optical communication in IMDD systems. Instead of assigning wavelength, polarization, and intensity to fixed, separately recovered functions, the unified resource-pool framework organizes them as a composite optical state space and recovers that space through an integrated disordered photonic processor. The central contribution is therefore not simply the demonstration of a larger symbol alphabet, but a communication architecture in which multidimensional optical resources are jointly defined, optically projected, and directly recovered.

This distinction separates the present approach from conventional multidimensional multiplexing. In standard IMDD systems, wavelength scaling typically requires additional multiplexing and demultiplexing branches, while polarization is usually restricted to two orthogonal channels with a matched receiver architecture [34]. Such schemes remain effective, but their capacity scaling is closely coupled to dimension-by-dimension system organization. Here, the integrated disordered processor does not separately recover wavelength, polarization, and intensity. Instead, it maps each composite optical state into a reduced set of output features, allowing enlarged state spaces to be recovered without assigning a dedicated receiver branch to each constituent optical variable. The use of multiple polarization states should therefore not be interpreted as increasing the intrinsic dimensionality of single-mode polarization; rather, polarization is reorganized from a binary channel into a multi-state component of a larger wavelength–polarization–intensity alphabet.

The experimental results support this architectural interpretation. The representative dual-wavelength implementation realizes a 4096-state composite alphabet corresponding to 12 bits per symbol slot, while dimensional-expansion experiments show that the same joint-projection principle can support denser polarization alphabets and wavelength-indexed state spaces. Fiber-link measurements further confirm that the projected fingerprints remain recoverable under loss, receiver-noise, and launch-power variations. In standard single-mode fiber, high-power operation is eventually limited by nonlinear and back-scattering impairments, with dimension-resolved analysis showing that the intensity component is most strongly affected. Hollow-core fiber mitigates this limitation by reducing light–glass interaction and nonlinear penalties, thereby extending the power margin for preserving distinguishable composite optical states[38,40]. In this sense, hollow-core fiber is not the basis of the framework, but a compatible transmission platform for testing its scalability under more demanding link conditions.

The projection principle also gives the framework a degree of generality beyond the specific communication experiment. Continuous wavelength drift and dynamic polarization fluctuations can alter the projected fingerprints and increase residual uncertainty, but the measurements in **Supplementary Note 8** show that the disordered projection retains robust and deterministic physical variants for stable recovery under these perturbations. More broadly, a calibrated photonic projection followed by a lightweight decision layer may be useful wherever multidimensional optical states must be classified, including optical-state sensing[41], photonic computing interfaces[42] and multimodal measurement systems[18]. These extensions should be viewed as implications of the projection concept, rather than as separate demonstrations in the present work.

The present work also has clear boundaries. The experiments demonstrate recoverability of enlarged composite optical state spaces, but do not alter the Shannon limit of the underlying optical channel. The gain lies in reorganizing available optical variables into a more efficiently usable direct-detection architecture, not in creating capacity beyond fundamental channel constraints. The current packaged prototype also remains a proof-of-concept system. Although the integrated photodetectors show 22.6-GHz optoelectronic bandwidth, the multi-channel package and readout electronics were not designed for a full high-baud resource-pool communication link. Further progress will require RF-optimized packaging, calibrated high-speed detector arrays, high-speed integrated polarization modulators, real-time decoding, insertion-loss reduction, fabrication-tolerance analysis, energy-per-bit evaluation and long-term environmental stabilization.

These results show that high-dimensional optical communication in IMDD systems need not remain constrained by dimension-partitioned architectures. By combining a unified resource-pool framework with an integrated disordered photonic processor, this work demonstrates that wavelength, polarization, and intensity can be jointly organized as a composite optical state space and recovered through compact optical-domain projection rather than through separate receiver branches. This shifts capacity scaling from the addition of independently resolved channels towards the co-design of optical alphabets, propagation media and integrated photonic processors. More broadly, the work suggests a route to hardware-efficient high-dimensional direct-detection systems in which existing degrees of freedom of light are not merely multiplexed, but jointly transformed into recoverable communication resources.

## Methods

### Device fabrication

The integrated processor is fabricated on a silicon-on-insulator wafer consisting of a 220-nm silicon device layer, a 2-μm buried oxide layer and a 750-μm silicon handle wafer, followed by a 3-μm $SiO_2$ upper cladding. Spot-size converter is used for efficient input coupling. The polarization splitting-rotator is realized with a 90-nm slab and an adiabatic bi-level taper, enabling polarization conversion and routing on chip. The disordered air-hole region is etched through the full 220-nm silicon layer. The integrated photodetectors are waveguide-coupled germanium–silicon (Ge–Si) p–i–n devices. The p-type regions ae defined in the silicon layer by multi-dose boron implantation, followed by the growth of a 500-nm-thick Ge absorption layer. Phosphorus ions are subsequently implanted into the upper Ge region to a depth of approximately 100 nm, thereby completing the vertical p–i–n junction.

**Experimental setup**

The experiment is conducted using the independent Tx and Rx ends. At the transmitter stage, two channels from a four-channel tunable laser source are selected as carriers. Each optical path is independently processed through an intensity modulator driven by a dual-channel arbitrary waveform generator. Subsequently the signals are modulated by a high-speed polarization scrambler to encode the 16-state polarization alphabet. The resulting multidimensional signals are combined and launched into a standard SMF link. At the receiver, a variable optical attenuator regulates the power level before the signal enters the integrated disordered photonic processor, with the final four-port output fingerprints captured by a high-speed real-time oscilloscope. A detailed schematic of the complete system link including every optical component is provided in **Supplementary Note 4** and illustrated in Figure S1.

**Signal Processing and neural network training**

The raw photocurrents undergo a tiered signal processing workflow across two software platforms. Initial preprocessing is conducted in MATLAB where the recorded waveforms are denoised and subjected to Z-score normalization to standardize the feature distribution. To further enhance the discrimination capability of the model, the four original spatial features are mathematically expanded into a sixteen-dimensional enhanced feature vector. These processed datasets are then imported into Python for the implementation of the neural network. The recovery of composite states is performed by a multi-task deep learning model utilizing depthwise separable convolutions to ensure computational efficiency. The model is trained on a high-performance workstation using a cross-entropy loss function and the Adam optimizer. The comprehensive data processing flowchart and the detailed topology of the neural network architecture are presented in **Supplementary Note 6** and Figure S3.

**Data Availability**

The data supporting the findings of this study are available within the paper and its Supplementary Information, and are also available from the corresponding authors upon request.

**Code Availability**

The open-source code is available in xxxx

**Acknowledgements**

This work is supported by the National Key Research and Development Program of China (No. 2025YFA1212900), the National Natural Science Foundation of China (No. 62225110，No. 62175079, and No. 62205119), Hubei Province Technological Innovation Program Project (2025BAA003) and Hubei Optical Fundamental Research Centre.

**Author contributions**

S.Y., Y.Y., and M.T., conceived and planned the research. Y.Y. and Z.G. designed and fabricated the devices; J.L., Z.G., X.Y., and Z.Z. designed and implemented the experimental transmission system and the machine learning algorithms realized by J.L. and J.Y.; J.L., Z.G., and X.Y. led the optical fiber measurements and data analysis with assistance from Z.Z., Z.L and Y.X.; J.L., X.Y., Z.G. wrote the paper with inputs from all the co-authors. The work was supervised by S.Y., Y.Y and M.T.

**Competing interests**

The authors declare no competing interests.

**Additional information**

**Supplementary information** The online version contains supplementary material available at https://doi.org/xxxxxx.